\documentclass[aps,preprint,showkeys]{revtex4-1}
\usepackage{graphicx}
\usepackage{amsmath}
\usepackage{makeidx}
\usepackage{amsfonts}
\usepackage{amssymb}
\usepackage{mathtools}
\usepackage{xcolor}

\begin{document}

\title{Efficiency of energy-consuming random walkers: Variability in energy helps}
\date{\today}

\author{Mohsen Ghasemi Nezhadhaghighi}
\affiliation{Department of Physics, Collage of Science, Shiraz University, Shiraz 71454, Iran}

\author{Abolfazl Ramezanpour}
\email{aramezanpour@gmail.com}
\affiliation{Department of Physics, Collage of Science, Shiraz University, Shiraz 71454, Iran}
\affiliation{Leiden Academic Centre for Drug Research, Faculty of Mathematics and Natural Sciences, Leiden University, PO Box 9500-2300 RA Leiden, The Netherlands}

\begin{abstract}
Energy considerations can significantly affect the behavior of a population of energy-consuming agents with limited energy budgets, for instance, in the movement process of people in a city. We consider a population of interacting agents with an initial energy budget walking on a graph according to an exploration and return (to home) strategy that is based on the current energy of the person. Each move reduces the available energy depending on the flow of movements and the strength of interactions, and the movement ends when an agent returns home with a negative energy. We observe that a uniform distribution of initial energy budgets results in a larger number of visited sites per consumed energy (efficiency) compared to case that all agents have the same initial energy if return to home is relevant from the beginning of the process. The uniform energy distribution also reduces the amount of uncertainties in the total travel times (entropy production) which is more pronounced when the strength of interactions and exploration play the relevant role in the movement process. That is variability in the energies can help to increase the efficiency and reduce the entropy production specially in presence of strong interactions.
\end{abstract}


\maketitle

\section{Introduction}\label{S1}

Human everyday activities (e.g., commuting between home and workplace, or shopping) are specific kinds of recurrent diffusion process in which agents travel from one place to another and then return to their starting point after following a trajectory with a number of steps of various lengths \cite{Gonzalez,Scafetta2011,Barbosa}. An exploration and return (ER) strategy is usually used to model such movements in social and ecological systems \cite{song-2010,Majumdar2010,Benichou2011,Pappalardo,wang2022}. This basic principle has been generalized in many studies to better describe for instance the social, economical, and geometrical aspects of these systems \cite{gonzalez-2015,vazifeh2021}.  

In the past few years, the emergence of location tracking devices (e.g., GPS navigator and smart phones), and location-based services (e.g., Foursquare, Yelp checkin and Google places) provides good opportunities to study human mobility patterns at very different spatial and temporal scales \cite{gallotti-2012,barthelemy-2019,bettencourt-2021,batty-2021,gonzalez-2022}: from mobility of individuals inside a city to mobility and transportation in an entire country \cite{Chowel,Brockmann,Vespignani,gonzalez-2023}. As a consequence, important progresses have been made from reconstruction of population density, mobility patterns and flows \cite{Phithakkitnukoon,Kitamura,Peng}, traffic forecasting and urban planning \cite{Nagel,Wang,Rozenfeld}, marketing campaign and prediction of epidemics \cite{Fibich,Pastor-Satorras}, to designing of mobile network protocols \cite{Chaintreau}.

Efficiency of structure and dynamical processes is essential to maintain a sustainable system like a city \cite{horner-2002,newman-2006,dong-2016,latora-2018,indaco-2019,barthelemy-2022}. However, the majority of research on this subject have focused on simulating the statistical characteristics of human mobility, such as displacement and gyration radius. We know that energy, used by human body or vehicle, plays a key role in transportation and other forms of social activities \cite{Kolbl,boyer-2009,Wang22}. It has been observed that the average travel times for different transportation modes (e.g. walking, cycling, bus, or car travel) are inversely proportional to the (physiological) energy consumption rates measured for the respective physical activities \cite{Kolbl}. Interestingly, when daily travel-time distributions of different transport modes are appropriately scaled, they turn out to have a universal functional relationship \cite{Kolbl,kolbl-2021}.

In this work we are going to investigate the effects of energy consumption and limited energy budgets on a measure of efficiency and entropy production in a movement process which is based on the ER strategy.
We see how constraints on energy budgets can influence the movements of these interacting and energy consuming agents, for instance resulting in a subdiffusion regime. In particular, we are interested in the efficiency of such a movement process and its relation to a measure of entropy or uncertainty production in such a process \cite{indaco-2020,indaco-2021}. 
We study the number of distinct locations visited by a person per the consumed energy. A greater number of demands are expected to be fulfilled during an urban exploration when a larger number of distinct sites are visited. A reasonable definition of efficiency should also take into account the energy consumed by the agent in this process. Additionally, we are interested in the uncertainty in the travel times which is generated by interactions between the agents. Smaller uncertainties in the travel times are expected to result in a better planning and therefore smaller dissipation.

The paper is organized as follows. We start in Sec. \ref{S20} with a definition of the model and the main parameters. In Sec. \ref{S21} we study an effective one-dimensional model and write a master equation in the continuum limit which can easily be generalized to higher dimensions. A naive mean-field approximation to the equation is presented and the results are compared with the exact solutions of the master equation in one dimension. Section \ref{S22} is devoted to the numerical simulations of the model in two dimensions. Concluding remarks are given in Sec. \ref{S3}.

\section{Results and Discussion}\label{S2}

\subsection{Model}\label{S20}
We consider $M$ agents $i=1\dots,M$ moving on a two-dimensional square lattice of size $N=L\times L$. The agents (or persons) start the walks at $t=0$ from the origins (or homes) $O_i$ which are distributed according to a reasonable distribution. We use the population growth model introduced in Ref. \cite{Li2017} to construct an initial distribution of the population. More precisely, we start from a unit of population at the center of the 2D square lattice. In each step, one site $a$ is chosen with a probability proportional to its population $M_a+c$ and a unit of population is added if site $a$ or at least one of its nearest neighbors is populated ($c$ is free parameter, which $c\ll1$ leads to the highly dense population near the center and $c\gg1$ leads to the homogeneous density of population all over the city. In the following we work with $c=1$. The process continues while the whole population $\sum_{a=1}^N M_a$ is less than $M$. This growth process generates spatially heterogeneous population densities which are qualitatively close to the empirical distributions.

Each person $i$ is given an initial energy budget $E_i(0)$ which is consumed in the links $l$ of the path taken by the person. 
We assume that the change in energy is proportional to a measure of the travel time in link $l$, i.e., $\Delta E_i=-\epsilon \tau_l(t)$ with a positive coefficient $\epsilon$. The latter depends on the flow $J_l(t)$ of movements in that link at time step $t$ when the person enters the link \cite{USB,transport-2011},
\begin{align}\label{tau_l}
\tau_l(t)=1+h\left(\frac{J_l(t)}{J_l^*}\right)^{\mu}.
\end{align}
The parameters $h$ and $\mu$ represent the strength of interactions, that is how the flows affect the energy cost given the line capacity $J_l^*$. The parameter $2<\mu<5$ according to the empirical observations \cite{USB}. The line capacity associated to the link $l$ is given by
\begin{eqnarray}\label{Jstar}
J_l^* = \rho(1+\kappa \frac{\rho_l}{\rho}),
\end{eqnarray}
where, $\rho = \frac{M}{N}$ representing the population density. The case with $\kappa=0$ represents a uniform line capacity all over the city, while $\kappa=1$ represents the case in which the line capacity is proportional to the local population density $\rho_l$. In the following, we set $\rho_l=(M_{a_l}+M_{b_l})/2$, i.e., equal to the average population of the end nodes of link $l$.

At $t=0$ the initial population $\{M_a(0):a=1,\dots,N\}$ is given by the population distribution generated by the growth model. The initial energies are sampled either from a delta distribution $P_0(E)=\delta_{E,\bar{E}}$ where all persons have the same energy budget or from a uniform distribution $P_0(E)\propto 1$ with a maximal variability in energy ($0<E<2\bar{E}$) but the same average energy $\bar{E}=\sum_E E P_0(E)=\langle E(0)\rangle$. The Kronecker delta function $\delta_{x,y}=1$ for $x=y$, otherwise it is zero. At each time step of size $\Delta t=1$, each person $i$ with current position $a$ and energy budget $E_i$ tries to change its position. With probability 
\begin{align}
\alpha(E_i)=\frac{1}{1+e^{-\beta E_i/\bar{E}}},
\end{align} 
the next position is chosen with equal probability from the set of nearest neighbors of $a$. Here, the positive parameter $\beta$ controls the probability of exploration depending on the available energy $E_i$. If the agent has enough energy, i.e., $E_i>\bar{E}$, then a large $\beta$ leads to a large exploration probability $\alpha(E_i) \sim 1$. 

With complementary probability $1-\alpha(E_i)$, the person goes to the nearest neighbor which is closer to its home $O_i$. 
In words, for positive $\beta$, the moves resemble a simple random walk (exploration is dominated) if the available energy $E_i$ is larger than the energy scale $\bar{E}$. On the other hand, the moves are mostly directed towards the origin (return is dominated) when the available energy is smaller than $-\bar{E}$. After any move the energy reduces by $\Delta E_i$ as described above. In numerical simulations, we shall end the movement if the person returns home with a negative energy. Moreover, the travel time is measured with the number of travels $t$; the waiting time $\tau_l(t)$ which is associated to link $l$ at time step $t$ just affects the dynamics by decreasing the available energy.  Note that for negative $\beta$, the probability of returning to the origin would be significantly larger than the exploration at the initial stages of the process. In the following, however, we shall consider the more realistic case of $\beta >0$.

\subsection{An effective model}\label{S21}
In this section we consider a single person walking on a one-dimensional chain. The effects of interaction with the others can be modeled by the way that energy is degraded. Let $\rho(x,E:t)$ be the probability of being at position $x$ and with energy $E$ at time $t$. Take $p(x,E)$ and $q(x,E)$ as the probabilities of moving to the right and left directions, respectively. The master equation governing $\rho(x,E:t)$ reads as follows, 
\begin{multline}\label{Deq}
\rho(x,E:t+\delta t)=(1-p(x,E)-q(x,E))\rho(x,E:t)\\
+p(x-\delta x,E+\delta E_R)\rho(x-\delta x,E+\delta E_R:t)+q(x+\delta x,E+\delta E_L)\rho(x+\delta x,E+\delta E_L:t),
\end{multline}
where, assuming a constant line capacity $J^*$, the changes in energy are
\begin{align}\label{dE}
\delta E_R=\epsilon(1+h(\frac{J_R}{J^*})^{\mu})\delta t,\\
\delta E_L=\epsilon(1+h(\frac{J_L}{J^*})^{\mu})\delta t,
\end{align}
depending on the right and left currents
\begin{align}
J_R(x:t)=\int dE p(x,E:t)\rho(x,E:t),\\
J_L(x:t)=\int dE q(x,E:t)\rho(x,E:t).
\end{align}

\begin{figure}
\includegraphics[width=16cm]{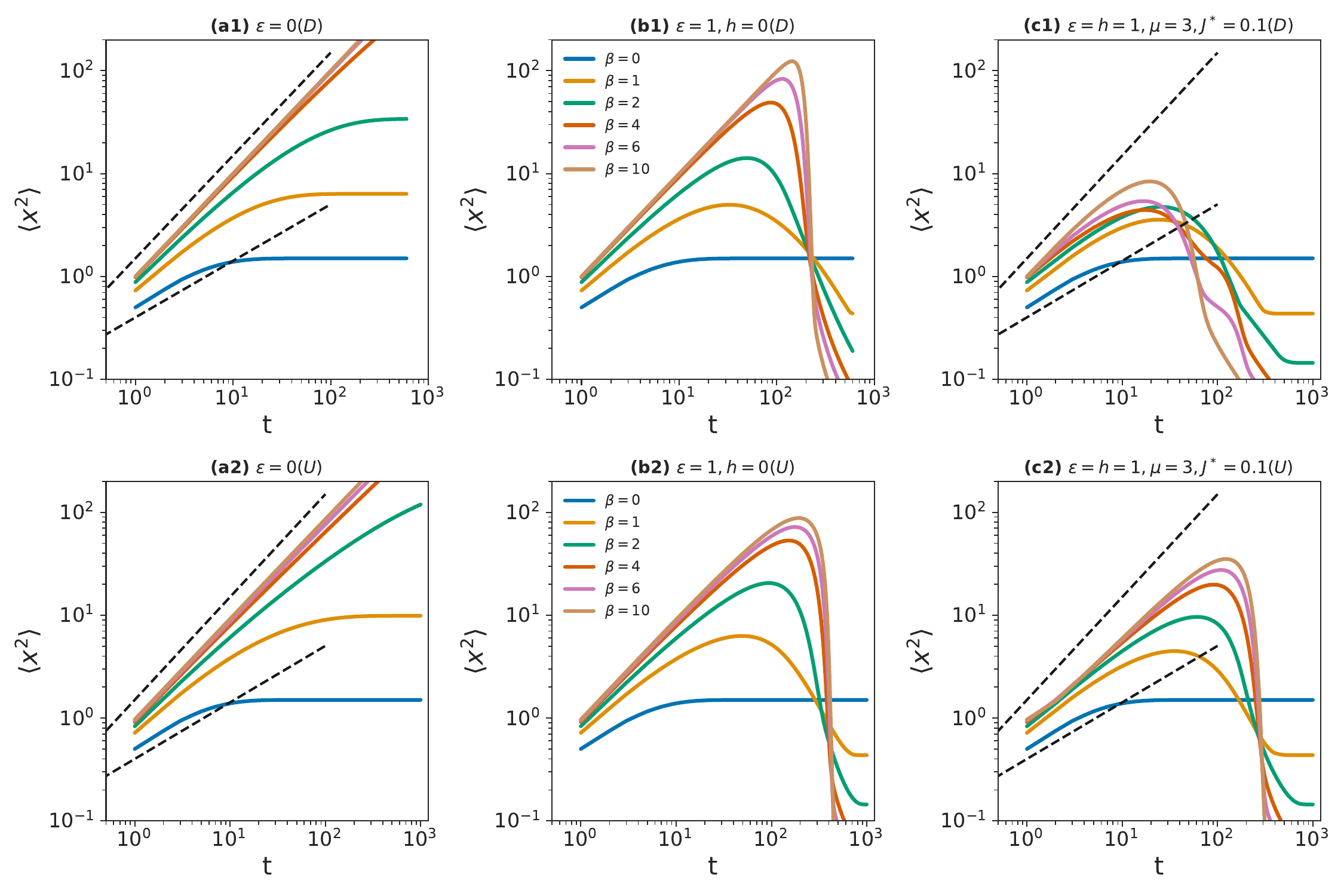} 
\caption{The mean of squared displacement $\langle x^2\rangle$ vs time step $t$ in the effective one-dimensional model. The results are obtained by numerical solution of the discrete master equation for a delta (panels (a1),(b1),(c1)) and uniform (panels (a2),(b2),(c2)) energy distribution. In panels (a1) and (a2) no energy is consumed $\epsilon=0$. In panels (b1) and (b2) the rate of energy consumption $\epsilon=1$ but there is no interaction $h=0$. In panels (c1) and (c2) the rate of energy consumption $\epsilon=1$ and interaction strength $h=1$ with $\mu=3, J^*=0.1$. The dashed lines show the $t$ and $t^{1/2}$ behaviors.}\label{f1}
\end{figure}

\begin{figure}
\includegraphics[width=10cm]{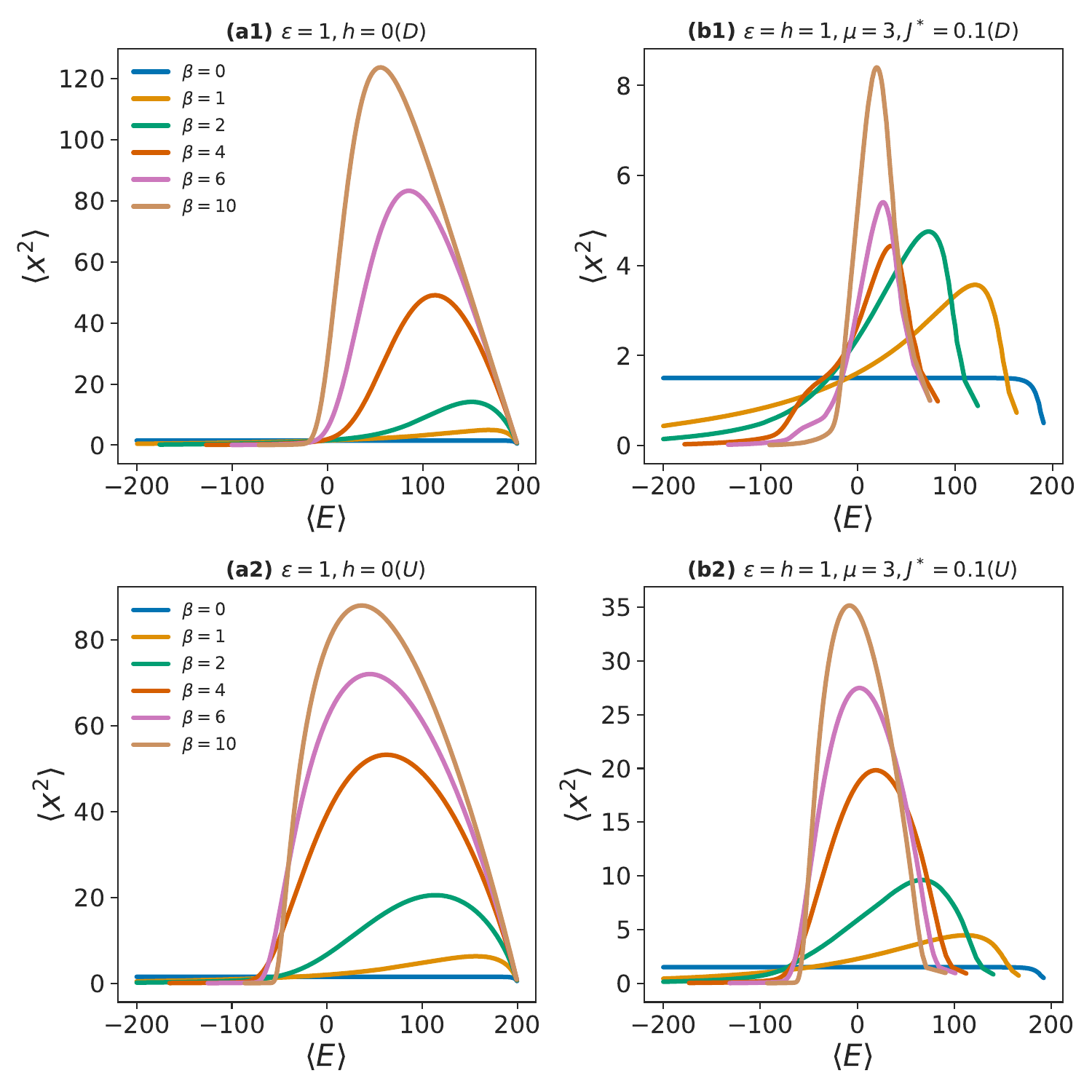} 
\caption{The mean of squared displacement $\langle x^2\rangle$ vs the current energy $\langle E\rangle$ in the effective one-dimensional model. The results are obtained by numerical solution of the discrete master equation for a delta (panels (a1),(b1)) and uniform (panels (a1),(b2)) energy distribution. In panels (a1) and (a2) the rate of energy consumption $\epsilon=1$ but there is no interaction $h=0$. In panels (b1) and (b2) the rate of energy consumption $\epsilon=1$ and interaction strength $h=1$ with $\mu=3, J^*=0.1$.}\label{f2}
\end{figure}

The above discrete equations (with $\delta t=\delta x=1$) are solved numerically to obtain the mean of squared displacement $\langle x^2(t)\rangle=\sum_{x,E} x^2\rho(x,E:t)$, where $\langle \dots \rangle$ denotes averaging with respect to $\rho(x,E:t)$. A particle of average energy $\bar{E}=200$ is released at time $t=0$ from the center of a chain of length $L=200$. The mean of squared displacement is reported in Figs. (\ref{f1}) and (\ref{f2}) vs time step $t$ and the average energy $\langle E(t)\rangle=\sum_{x,E} E\rho(x,E:t)$. In the following, we may omit the time dependence of the average quantities, for instance  using $\langle E\rangle$ for $\langle E(t)\rangle$. Note that here we are solving Eq. \ref{Deq} numerically for the probability distribution $\rho(x,E:t)$ starting from an initial condition $\rho(x,E:0)=\delta_{x,L/2}\delta_{E,\bar{E}}$. We observe that $\langle x^2\rangle$ is initially proportional to $t$ (normal diffusion) or $t^{1/2}$ (subdiffusion) for very large or small $\beta$, respectively. That is $\langle x^2\rangle\propto t^{\nu}$ for not very large $t$ with $1/2\le \nu\le 1$. The average number of distinct visited sites in this one-dimensional problem can be estimated as $2\sqrt{\langle x^2\rangle}$. Then, from Fig. \ref{f1} (panels (c1),(c2)) and Fig. \ref{f2} (panels (b1),(b2)), we observe that this number is significantly larger in presence of interactions ($h>0$) when the initial energies are distributed uniformly. Later, in Sec. \ref{S22} we will compare the number of distinct visited sites with the consumed energy to have a better measure of efficiency in these processes.

\subsubsection{The continuum limit}\label{S211}
Here we write the master equation in the continuum limit and present a mean-field solution to the equations. It would be then straightforward to generalize the resulted differential equation to higher dimensions.  
In the continuum limit $\delta t, \delta x \to 0$, the main equation for the probability density $\rho(x,E:t)$ of position and energy of the walker at time $t$ can be written as follows
\begin{multline}
\frac{\partial}{\partial t}\rho(x,E:t)=\frac{\partial^2}{\partial x^2}(D(x,E)\rho(x,E:t))
-\frac{\partial}{\partial x}(f(x,E)\rho(x,E:t))\\+2r(x:t)\frac{\partial}{\partial E}(D(x,E)\rho(x,E:t)), 
\end{multline}
where we defined
\begin{align}
D(x,E) &=\frac{(\delta x)^2}{2\delta t}(p(x,E)+q(x,E)),\\
f(x,E) &=\frac{\delta x}{\delta t}(p(x,E)-q(x,E)),\\
r(x:t)&=\frac{\delta E}{\delta t},
\end{align}
where $D(x,E)$ represents the diffusion coefficient, $f(x,E)$ defines drift (external force), and $r(x,E)$ is the rate of energy consumption.
The diffusion coefficient $D(x,E)$ represents the stochastic and unbiased part of the motion whereas the drift function $f(x,E)$ quantifies the effect of biases in the process. In general, these functions could depend on the current position and energy of the particle.

Here only terms of order $\delta t$ are taken; in the following we assume that $\delta E\propto \delta t$ and $\delta x\propto \sqrt{\delta t}$ and we set $(\delta x)^2/(\delta t)=1$. Moreover, we assume that the right and left transition probabilities differ in a way that results in appropriate diffusion and drift functions,   
\begin{align}
p(x,E)&=\frac{d(x,E)}{2}+(1-d(x,E))\theta(-x)\sqrt{\delta t},\\
q(x,E) &=\frac{d(x,E)}{2}+(1-d(x,E))\theta(x)\sqrt{\delta t},
\end{align}
where $0\le d(x,E)\le 1$, should be an increasing function of $E$. As before we take
\begin{align}
d(x,E)=\frac{1}{1+e^{-\beta E/\bar{E}}},
\end{align} 
and assume that it only depends on energy but not the position of the particle. The scaling with $\sqrt{\delta t}$ is chosen to have a well-defined continuum limit for the drift function $f(x,E)$. The step function $\theta(x)=1$ for $x>0$, otherwise it is zero. 
In this way, the diffusion and drift coefficients are given by
\begin{align}
D(x,E) &=\frac{d(x,E)}{2},\\
f(x,E) &=(\theta(-x)-\theta(x))(1-d(x,E)).
\end{align}
Note that in the continuum limit $\delta E_R=\delta E_L=\delta E$, with
\begin{align}
\delta E &=\epsilon(1+h(\frac{J}{J^*})^{\mu})\delta t,\\
J(x:t) &=\int dE D(x,E)\rho(x,E:t).
\end{align}
The function $\delta E/\delta t$ determines the system dynamics and shows how the energy is dissipated; if interactions are negligible, the rate of dissipation is expected to be a constant, i.e., $\delta E/\delta t=\epsilon$. On the other hand, in presence of interactions we may have a dissipation rate that depends on $\rho(x,E:t)$. 
As before, the parameters $h, J^*$ and $\mu$ control the strength of interactions. 

Now from the master equation we can see how the averages $\langle x^2\rangle$ and $\langle E\rangle$ change with time:
\begin{align}\label{dtx2}
\frac{\partial}{\partial t}\langle x^2\rangle=\langle d(E)\rangle+2\langle xf(x,E)\rangle,
\end{align} 
and
\begin{align}\label{dtE}
\frac{\partial}{\partial t}\langle E\rangle=-\langle r(x:t)d(E)\rangle.
\end{align} 
We assumed in derivation that $x\rho(x,E:t)$ and $x^2\partial_x \rho(x,E:t)$ vanish for $x=\pm \infty$. In addition, we assume that $\rho(x,E:t)$ vanishes for $E=\pm \infty$. Recall that
\begin{align}
r(x:t) &=\epsilon(1+h(\frac{J}{J^*})^{\mu}),\\
J(x:t) &=\int dE \frac{d(E)}{2}\rho(x,E:t).
\end{align}

\subsubsection{A naive mean-field approximation}\label{S212}
\begin{figure}
\includegraphics[width=16cm]{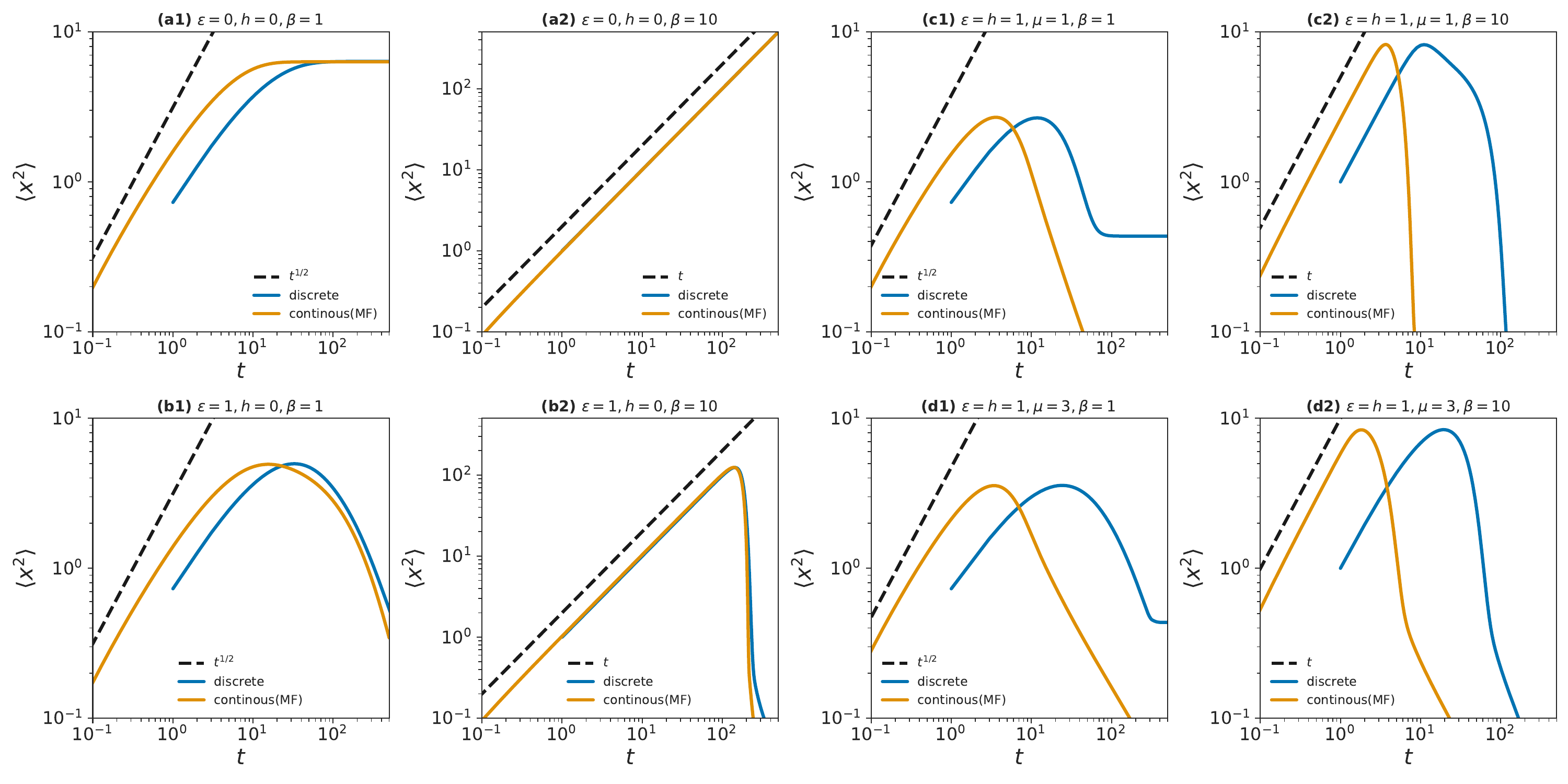} 
\caption{Comparing the mean-field results of the continuum model with the exact results of the discrete model. The same qualitatively behaviors is observed for different ranges of the parameters. The initial energy distribution in the discrete model is concentrated at $\bar{E}=200$ (a delta distribution) like the initial energy of the particle in the mean-field approximation. The rate of dissipation and strength of interaction are denoted by $\epsilon$ and $h$, respectively. In presence of interaction we set $J^*=0.1$. The dashed lines show the $t$ and $t^{1/2}$ behaviors.}\label{f3}
\end{figure}

We are going to approximate the average of products with the product of averages in equations \ref{dtx2} and \ref{dtE}. Note also that $\langle xf(x,E)\rangle =-\langle |x|(1-d(E))\rangle$ and  $J(x:t)=\langle \frac{d(E)}{2}\rangle$.
Then we get
\begin{align}
\frac{\partial}{\partial t}\langle x^2\rangle &\approx \langle d(E)\rangle-2\langle |x|\rangle(1-\langle d(E)\rangle),\\
\frac{\partial}{\partial t}\langle E\rangle &\approx -\langle r(x:t)\rangle \langle d(E)\rangle,\\
\langle r(x:t)\rangle &\approx \epsilon\left(1+h[\frac{\langle d(E)\rangle}{2J^*}]^{\mu}\right),\\
\langle d(E)\rangle &\approx \frac{1}{1+e^{-\beta\langle E\rangle/\bar{E}}}.
\end{align} 
We may also replace $\langle |x|\rangle\approx \sqrt{\langle x^2\rangle}$. In Fig. (\ref{f3}), we compare the above mean-field approximation with the exact results of the discrete model. Here, the initial energy distribution in the discrete model is concentrated at $\bar{E}$ (delta distribution) as the initial average energy of the particle in the mean-field approximation. 
We observe that the qualitative behavior of the discrete model is captured by the naive mean-field approximation of the continuum model for different parameter values. The approximation works well especially for small $h$ and large $\beta$. 

Note that for small times $t$, where the average energy is still positive, we have $d(E)=\frac{1}{2}, 1$ for $\beta=0, \infty$, respectively. Therefore, the second term with negative sign in $\frac{\partial}{\partial t}\langle x^2\rangle$ is more important when $\beta$ is small; for large $\beta$ it is a product of two small quantities $|x|$ and $1-d(E)$. Thus, we observe a subdiffusion for $\beta \to 0$, whereas for $\beta\to \infty$ a normal diffusion is observed for small times where the first term dominates. Nevertheless, the average energy decreases faster for larger $\beta$ than for smaller $\beta$. This for instance can be seen when $h=0$. In the absence of interactions, the average energy is given by
\begin{align}\label{Eh0}
\langle E\rangle-\bar{E}=\frac{\bar{E}}{\beta}(e^{-\beta \langle E\rangle/\bar{E}}-e^{-\beta})-\epsilon t,
\end{align}
with initial value $\langle E\rangle=\bar{E}$ at $t=0$. The time needed to reach average energy zero is 
\begin{align}
t(\langle E\rangle=0)=\frac{\bar{E}}{\epsilon}[1+\frac{(1-e^{-\beta})}{\beta}],
\end{align} 
which is a decreasing function of $\beta$. The relevant time scales in the mean-field equations are 
\begin{align}
\tau_0 &=\frac{1}{(1-\langle d(E)\rangle)^2},\\
\tau_{\epsilon} &=\frac{\bar{E}}{\epsilon\beta},
\end{align}
besides, the microscopic time $\delta t$, which is determined by the microscopic length $\delta x^2$. Recall that we set $\lambda=\frac{\delta x^2}{\delta t}=1$. In the absence of dissipation, energy is fixed and $\tau_0$ gives the time we need to reach the stationary state, which is finite as long as $d(E)<1$. On the other hand, for nonzero $\epsilon$ the time scale $\tau_{\epsilon}$ is the time we need to see the effects of dissipation in the energy. Note that dissipation results in a spectrum of time scales $\tau_0$ for different energies and interactions introduce a range of effective $\epsilon$ for different times.

Finally, we can say something about the scaling of $\beta$ to have an extended solution to the equations. In the stationary state, or for the maximum range $R=\sqrt{\langle x^2\rangle}$, we get
\begin{align}
\frac{\partial}{\partial t}\langle x^2\rangle &= \langle d(E)\rangle-2\langle |x|\rangle(1-\langle d(E)\rangle)=0,\\
\langle |x|\rangle &=\frac{\langle d(E)\rangle}{2(1-\langle d(E)\rangle)},
\end{align}
or
\begin{align}
R =\frac{1}{2}e^{\beta E_R/\bar{E}},
\end{align}
where $E_R$ is stationary energy or the average energy at the maximum range. Assuming that $E_R$ is on order of $\bar{E}$, the above relation says that to have an extended state $R \propto L$ we need $\beta \propto \ln L$, where $L$ is the linear size of system.

\subsection{Numerical simulations}\label{S22}

In this section we report the results of numerical simulations of the model described in Sec. \ref{S20} on a two-dimensional square lattice of size $N=L\times L$. To simplify the picture, in the following we set $\epsilon=1$ (i.e., energy is dissipated) and fix the parameter $\mu=3$, which is in the range of empirical values $\mu \in (2,5)$ \cite{USB}. The results are reported for two cases of $\kappa=0,1$ but we mainly focus on case $\kappa=1$ where the link capacities change with the local population (see Eq. \ref{Jstar}). The results vary more smoothly with the strength of interactions $h$ and are less sensitive to the size $N$ and population $M$ of the system for $\kappa=1$ compared to the case of uniform line capacity with $\kappa=0$. We set the initial average energy $\bar{E}=\langle E(0)\rangle=L^2$, to have enough energy for exploring the system by a simple Brownian motion.

Let us begin with the mean energy and the mean-squared-displacement (MSD) of agents, which are defined in terms of an average over the population
\begin{equation}
\langle E\rangle = \langle\frac{1}{M}\sum_{i=0}^M E_i(t)\rangle,
\end{equation}
\begin{equation}
\langle R^2 \rangle = \langle \frac{1}{M} \sum_{i=0}^M |\vec{r}_i(t)-\vec{r}_i(0)|^2\rangle.
\end{equation}

\begin{figure}
\includegraphics[width=16cm]{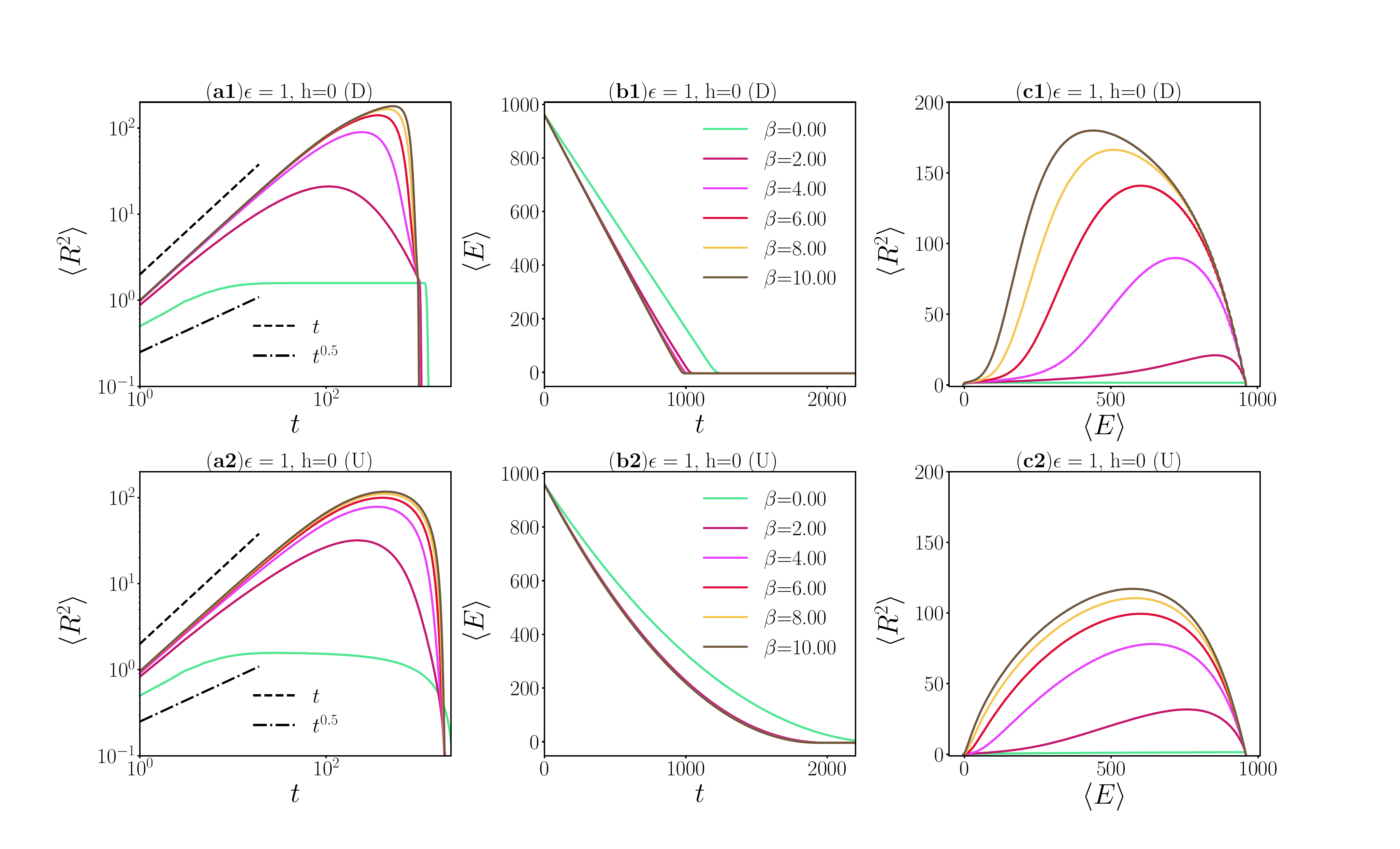} 
\caption{The mean of squared displacement $\langle R^2\rangle$ and the mean energy $\langle{E}\rangle$ versus time step $t$ for energy-consuming random walkers on the two-dimensional lattice of linear size $L=30$. The results are obtained by averaging over $500$ independent runs of the simulation. The top (bottom) panels show the results for the delta (uniform) energy distribution.  
The rate of energy consumption $\epsilon=1$ but there is no interaction $h=0$, and $\kappa=0$. The dashed lines in the left panels show the $t$ and $t^{1/2}$ behaviors.}\label{f4}
\end{figure}

\begin{figure}
\includegraphics[width=16cm]{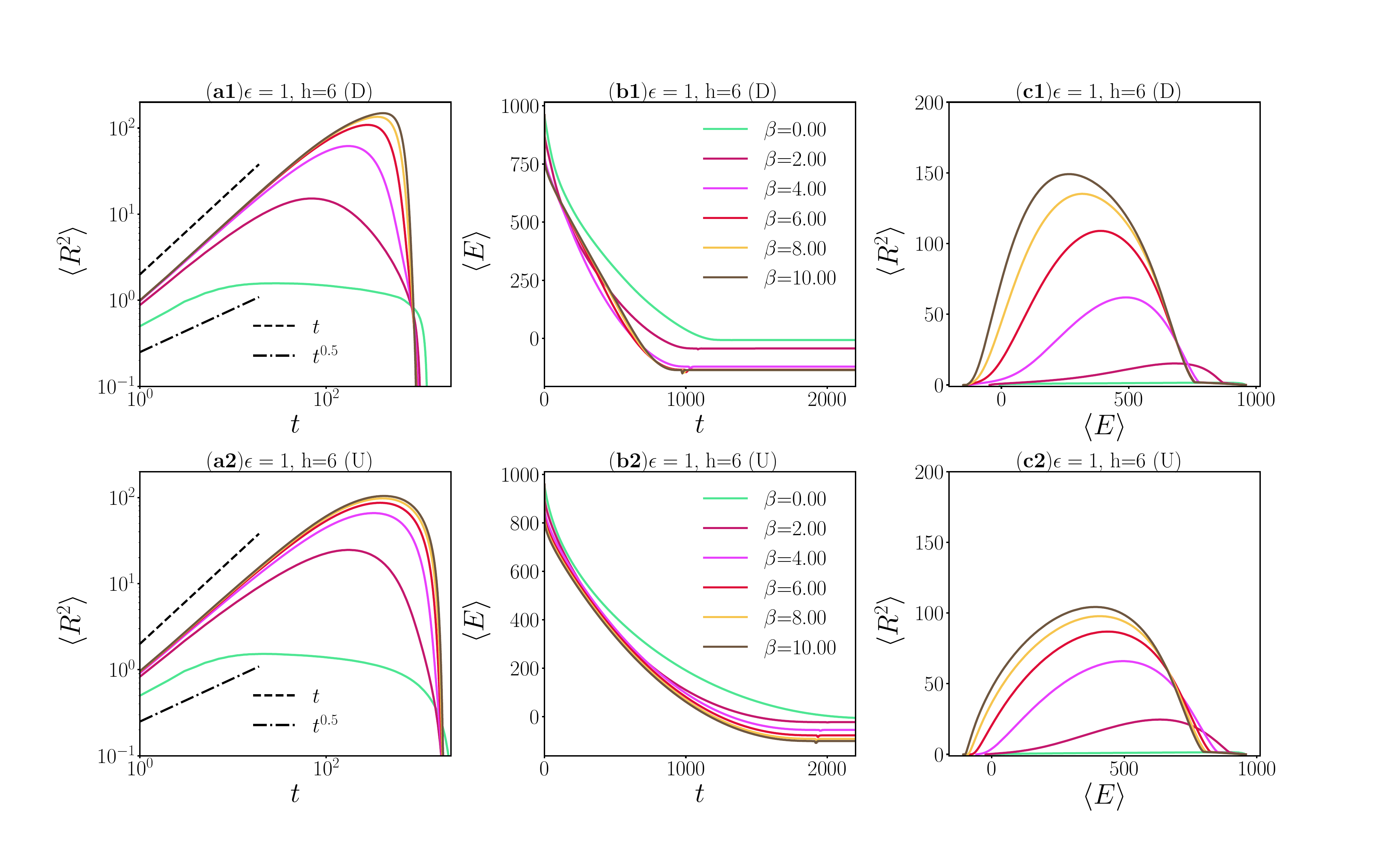} 
\caption{The mean of squared displacement $\langle R^2\rangle$ and the mean energy $\langle{E}\rangle$ versus time step $t$ for energy-consuming random walkers on the two-dimensional lattice with size $L=30$ . The results are obtained by averaging over $500$ independent runs of the simulation.  
The top (bottom) panels show the results for the delta (uniform) energy distribution.  
Here, the rate of energy consumption $\epsilon=1$ with interaction parameter $h=6$, and $\kappa=0$. The dashed lines in the left panels show the $t$ and $t^{1/2}$ behaviors.}\label{f5}
\end{figure}

Here $\langle \cdots \rangle$ denotes averaging over independent realizations of the movement process. Figures (\ref{f4}) and (\ref{f5}) show the MSD and $\langle{E}\rangle$ versus time from simulations for a range of values of the control parameters $h$ and $\beta$ (with $\kappa=0$). Note that for $\beta\gg 1$ the MSD grows linearly with time at the beginning where pure exploration dominants for each agents. In this regime the probability $\alpha(E)=\frac{1}{1+e^{-\beta E/\bar{E}}}$ is close to unity for any $E>0$ and the agent behaves like a simple random walker. On the other hand, for $h\to 0$ energy dissipation is independent of the traffic, thus after a time of order $\langle{E}(0)\rangle$, almost all the agents have negative energy. Then, return to the origin dominates and the MSD decreases rapidly to zero for $\beta\gg 1$. Figure \ref{f4} (b2) also shows that even in the absence of interaction, the average energy displays a nonlinear behavior when the initial energy distribution is uniform. That is because the initial energies are very different in this case and $\langle{E}\rangle$ is average of the total energy per person.

In the other side, we have the limit $\beta \to 0$ where return to origin is dominated and therefore $\langle R^2(t)\rangle$ grows sub-linearly with time. In this case a single trajectory of an agent belongs to the class of continuous-time random walks, where the agent's motion is interrupted by random waiting times. Mostly such processes are characterized by a scaling form of the mean squared displacement,
\begin{eqnarray}
\langle R^2(t)\rangle \sim t^{\nu},
\end{eqnarray} 
with $\nu<1$ \cite{metzler2000}. In our model we found that the scaling exponent $1/2<\nu <1$. As Figs. \ref{f4} and \ref{f5} show the average energy reduces more slowly when the initial energy has a uniform distribution. In contrast to the one-dimensional model of Sec. \ref{S21} here the MSD is usually larger for the delta energy distribution than the uniform distribution.

Next, to characterize more precisely the performance of the movement process, we study some reasonable measures of efficiency ${\eta}$ and entropy (uncertainty or disorder) production $S$ in the process. In particular, it would be interesting to find the parameters which result in higher efficiencies and lower entropy productions. The definition of efficiency ${\eta}$ here is based on the mean number of distinct sites visited by the agents and the consumed energy,
\begin{eqnarray}
\eta = \langle \frac{1}{M} \sum_{i=1}^M \frac{V_i}{\Delta E_i} \rangle. 
\end{eqnarray} 
Here $V_i$ is the number of distinct sites visited by agent $i$ and the denominator $\Delta E_i =E_i(0) - E_i(T_i)$ represents the total energy consumed by the agent. As mentioned before, an agent starts its movement by the initial energy $E_i(0)$ and stops when returns home with a negative energy $E_i(T_i)$ at time step $t=T_i$.

The average number of distinct sites $\langle V(t) \rangle $ visited by a random walker after $t$ steps has been well studied in the literature, for instance in \cite{Dvor}. For a foraging animal, $\langle V(t) \rangle $ corresponds to the size of the territory covered by the animal. In a search algorithm, this quantity corresponds to the number of configurations in the solution space that have been explored. The number of distinct locations visited by a person with a given amount of energy could be a good measure of efficiency of the process; for instance, visiting a larger number of distinct locations enhances the chance of satisfying a larger number of demands. 

\begin{figure}
\includegraphics[width=16cm]{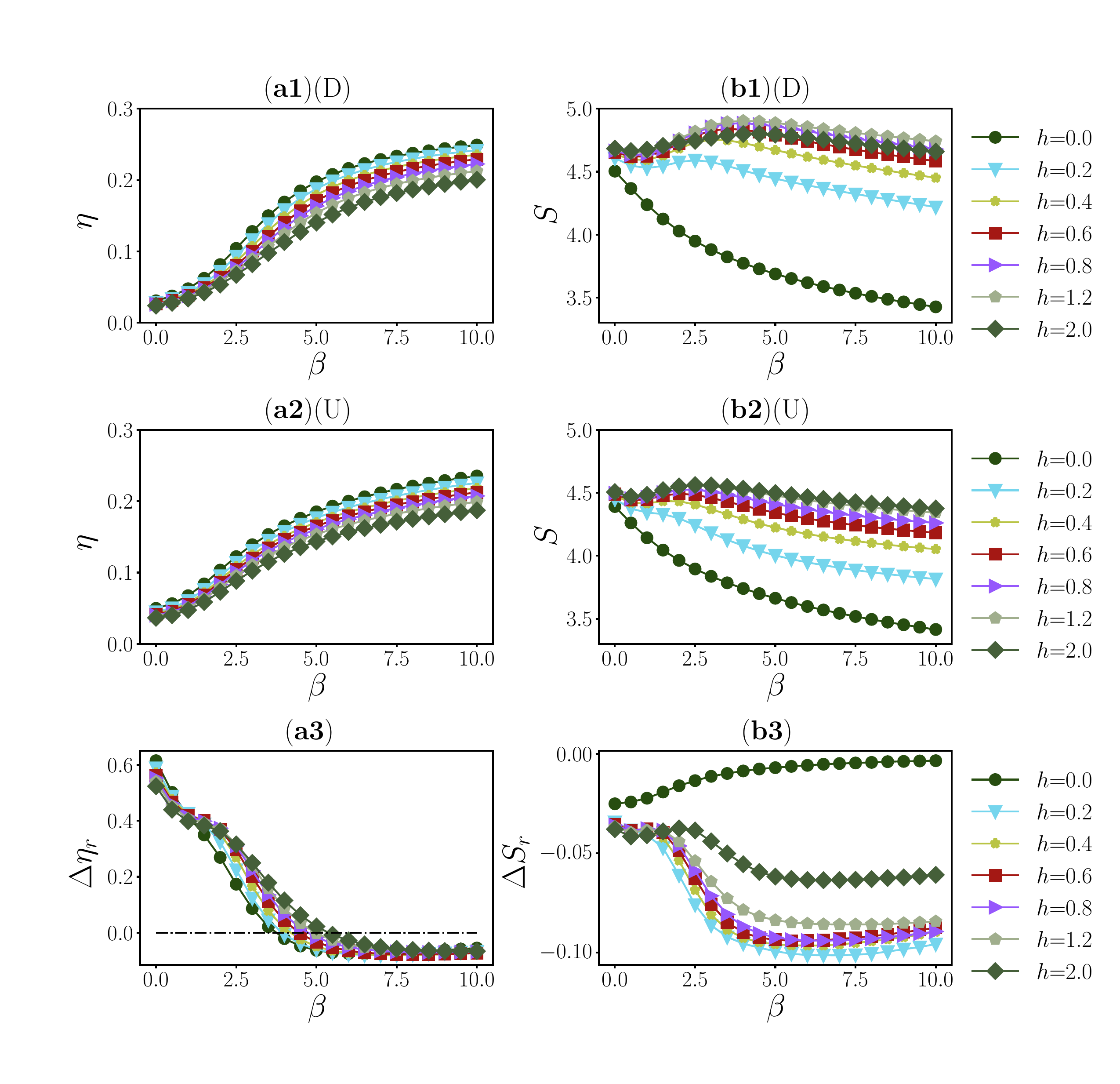} 
\caption{
Left: The efficiency $\eta$ vs $\beta$ for different interaction strength $h=\{0,0.2,0.4,0.6,0.8,1.2,2\}$.  
Right: The entropy production $S$ over the control parameter $\beta$ for different values of interaction strength $h$. 
The results are obtained by averaging over $200$ independent runs of simulation with lattice size $L=40$, $c=1.$ and $\kappa=0$.
Panels ($a_1$) and ($b_1$) show the results for the delta energy distribution and panels ($a_2$) and ($b_2$) show the results for the uniform energy distribution.
Panels ($a_3$) and ($b_3$) represents the relative differences $\Delta \eta_r = (\eta_U - \eta_D)/\eta_D$  and $\Delta S_r = (S_U - S_D)/S_D$ respectively. 
}\label{f6}
\end{figure}

To see how much the movement process increases disorder in the system we measure the amount of uncertainty in the return times of the agents. Consider a specific realization of the initial population distribution $M_a(0)$ and the initial energy distribution $E_i(0)$. We simulate the stochastic process of movement for $\mathcal{N}$ times to find an estimation of the probability distribution of the return times $\mathcal{P}_i(T_i)$. The entropy production of the process is then defined as:
\begin{eqnarray}
S = \frac{1}{M} \sum_{i=1}^M  S_i=\frac{1}{M} \sum_{i=1}^M \left(-\sum_{T_i}\mathcal{P}_i(T_i)\log[\mathcal{P}_i(T_i)]\right).
\end{eqnarray} 
   
\begin{figure}
\includegraphics[width=16cm]{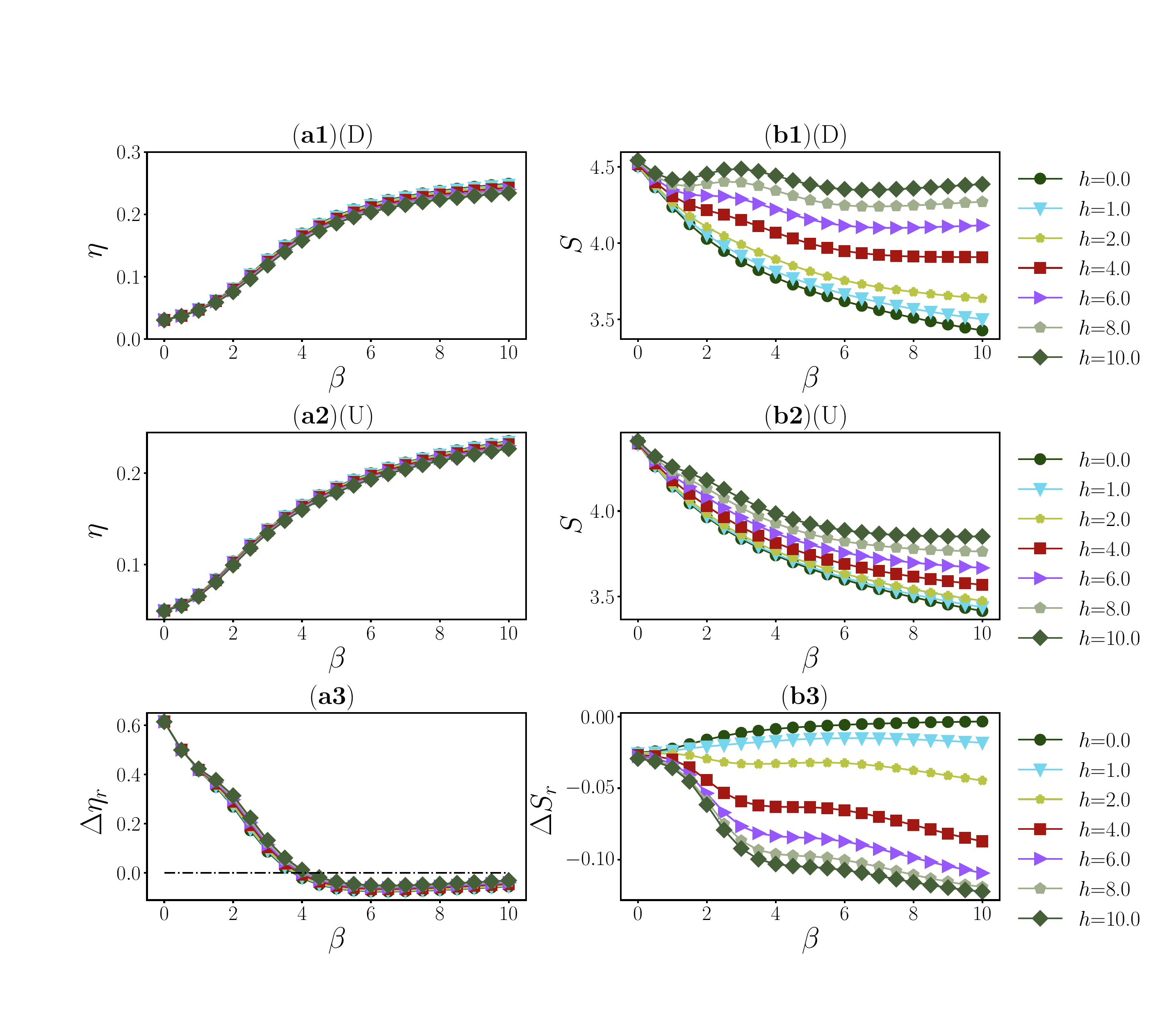} 
\caption{
Left: The efficiency $\eta$ vs $\beta$ for different interaction strength $h=\{0,1,2,4,6,8,10\}$.  
Right: The entropy production $S$ over the control parameter $\beta$ for different interaction strength $h$. 
The results are obtained by averaging over $200$ independent runs of simulation with lattice size $L=40$, $c=1.0$ and $\kappa=1$. Panels ($a_1$) and ($b_1$) show the results for the delta energy distribution (D) and panels ($a_2$) and ($b_2$) show the results for the uniform energy distribution (U). Panels ($a_3$) and ($b_3$) represents the relative differences $\Delta \eta_r = (\eta_U - \eta_D)/\eta_D$  and $\Delta S_r = (S_U - S_D)/S_D$ respectively. 
}\label{f7}
\end{figure}

\begin{figure}
\includegraphics[width=16cm]{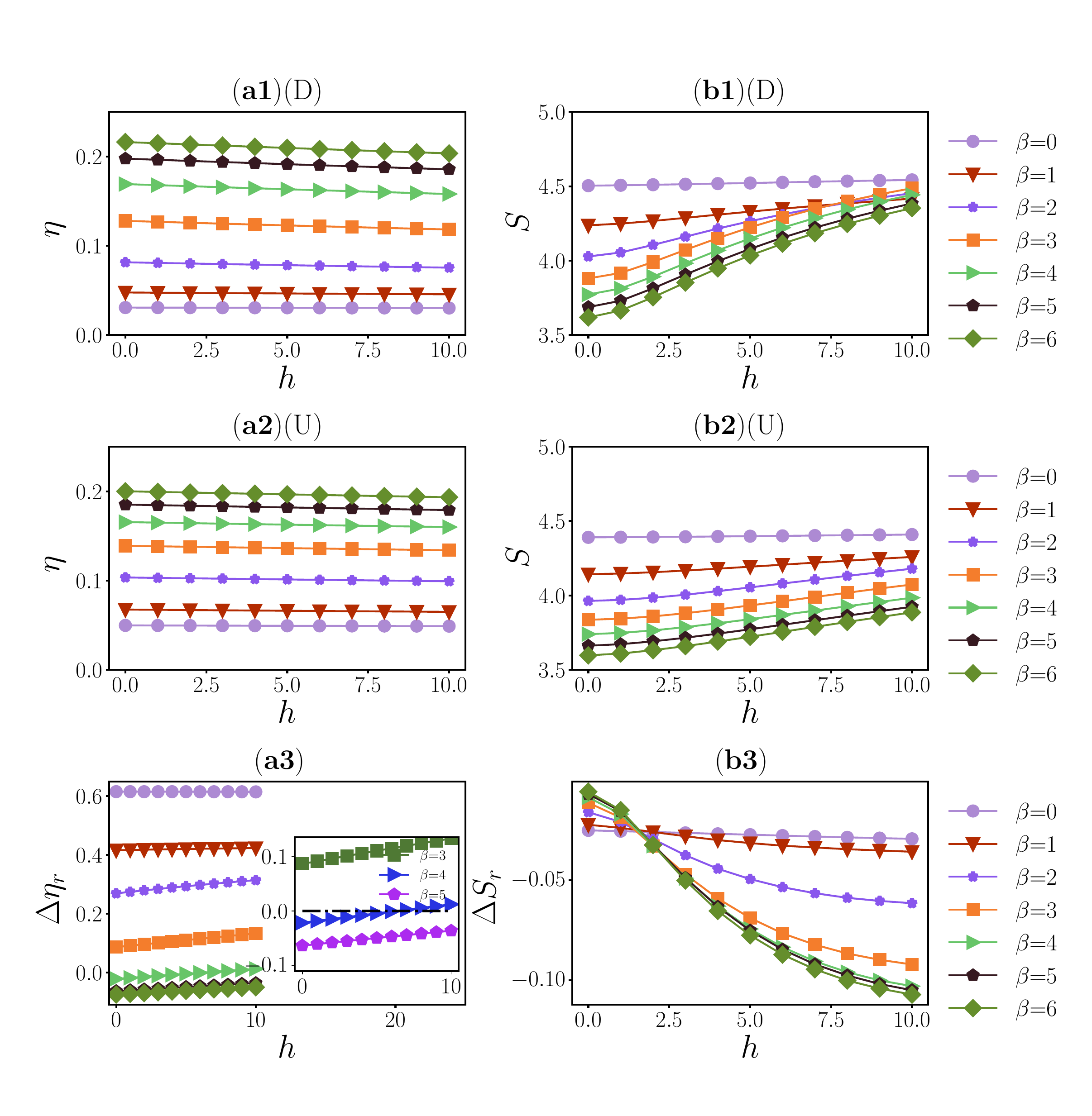} 
\caption{
Left: The efficiency $\eta$ vs interaction strength $h$ for different $\beta=\{0,1,2,3,4,5,6\}$.  
Right: The entropy production $S$ over the interaction strength $h$ for different values of $\beta$.  
The results are obtained by averaging over $200$ independent runs of simulation with lattice size $L=40$, $c=1.0$ and $\kappa=1$. 
Panels ($a_1$) and ($b_1$) show the results for the delta energy distribution (D) and panels ($a_2$) and ($b_2$) show the results for the uniform energy distribution (U).
Panels ($a_3$) and ($b_3$) represents the relative differences $\Delta \eta_r = (\eta_U - \eta_D)/\eta_D$  and $\Delta S_r = (S_U - S_D)/S_D$ respectively.  
}\label{f8}
\end{figure}

Figure (\ref{f6}) displays the efficiency $\eta$ and entropy $S$ for different values of $\beta$ and $h$ when the link capacities are constant, that is $\kappa=0$. Obviously the efficiency is reduced by introducing the interactions for a fixed exploration $\beta$. When interactions are negligible the number of visited sites plays the main role in the efficiency, thus the efficiency is an increasing function of $\beta$. For larger values of interactions, the consumed energy has the main contribution to the efficiency which could result in saturation or even reduction of the efficiency for very large $\beta$. The entropy of return times changes from a decreasing to and increasing function of $\beta$ as the strength of interactions $h$ gets larger than about $0.2$(for delta distribution) and $0.4$(for uniform distribution). It is in this region that equation $S(\beta)=s_0$ is satisfied for multiple values of $\beta$. In words, there are movement processes which have the same uncertainty $S$ but different efficiencies $\eta$. Moreover, we observe that except for very large $\beta$ and small $h$, the uniform distribution of energy budgets performs better than the delta distribution in both the efficiency and the generated uncertainty in the return times.

\begin{figure}
\includegraphics[width=16cm]{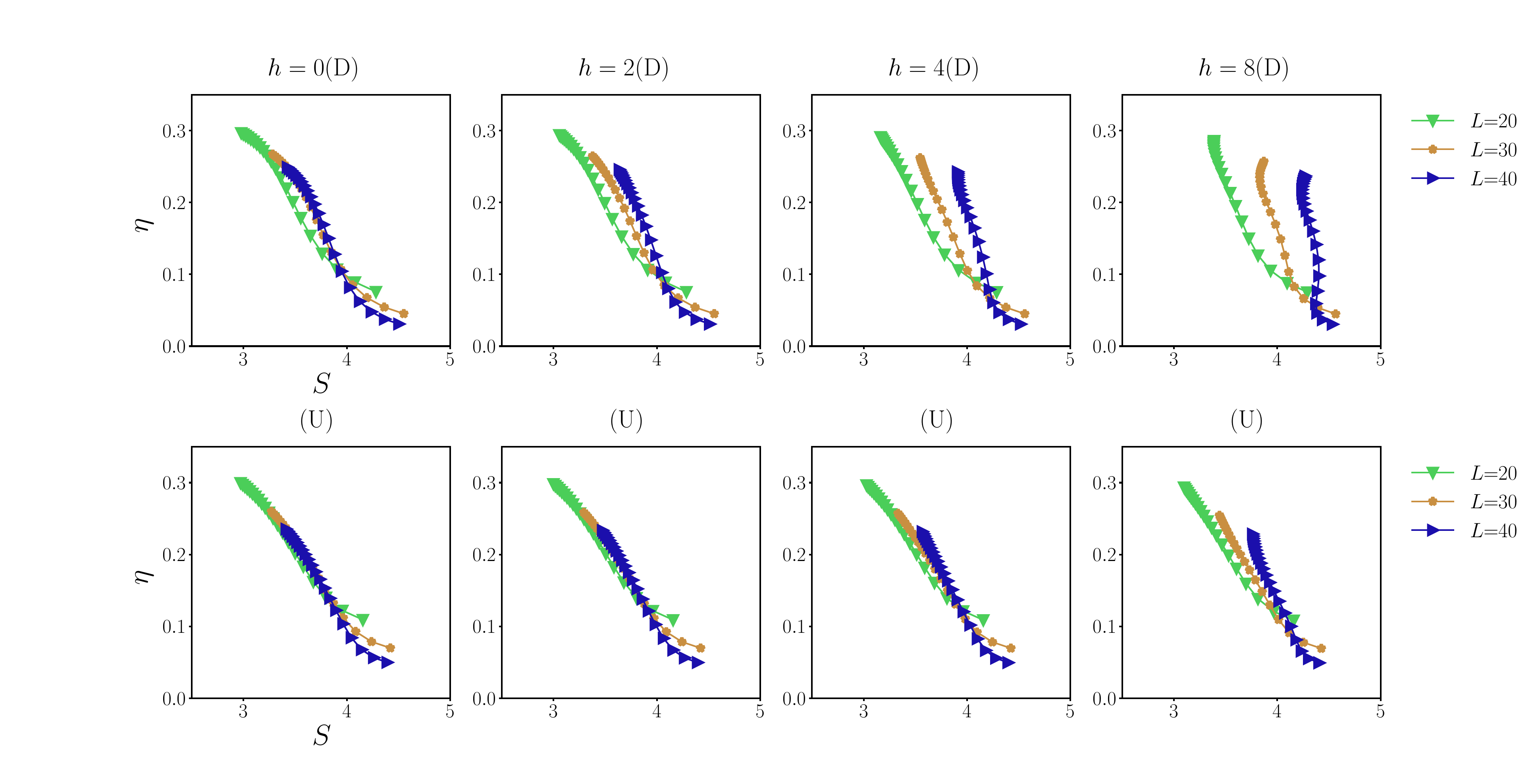} 
\caption{
The efficiency $\eta(\beta)$ vs the entropy production $S(\beta)$ for different $h=\{0,2,4,8\}$. The results are obtained by averaging over $1000,500,200$ independent runs of simulation with lattice sizes $L=20,30,40$, $c=1.0$ and $\kappa=1$. Top (bottom) panels show the results for the delta (uniform) energy distribution.  
}\label{f9}
\end{figure}

\begin{figure}
\includegraphics[width=16cm]{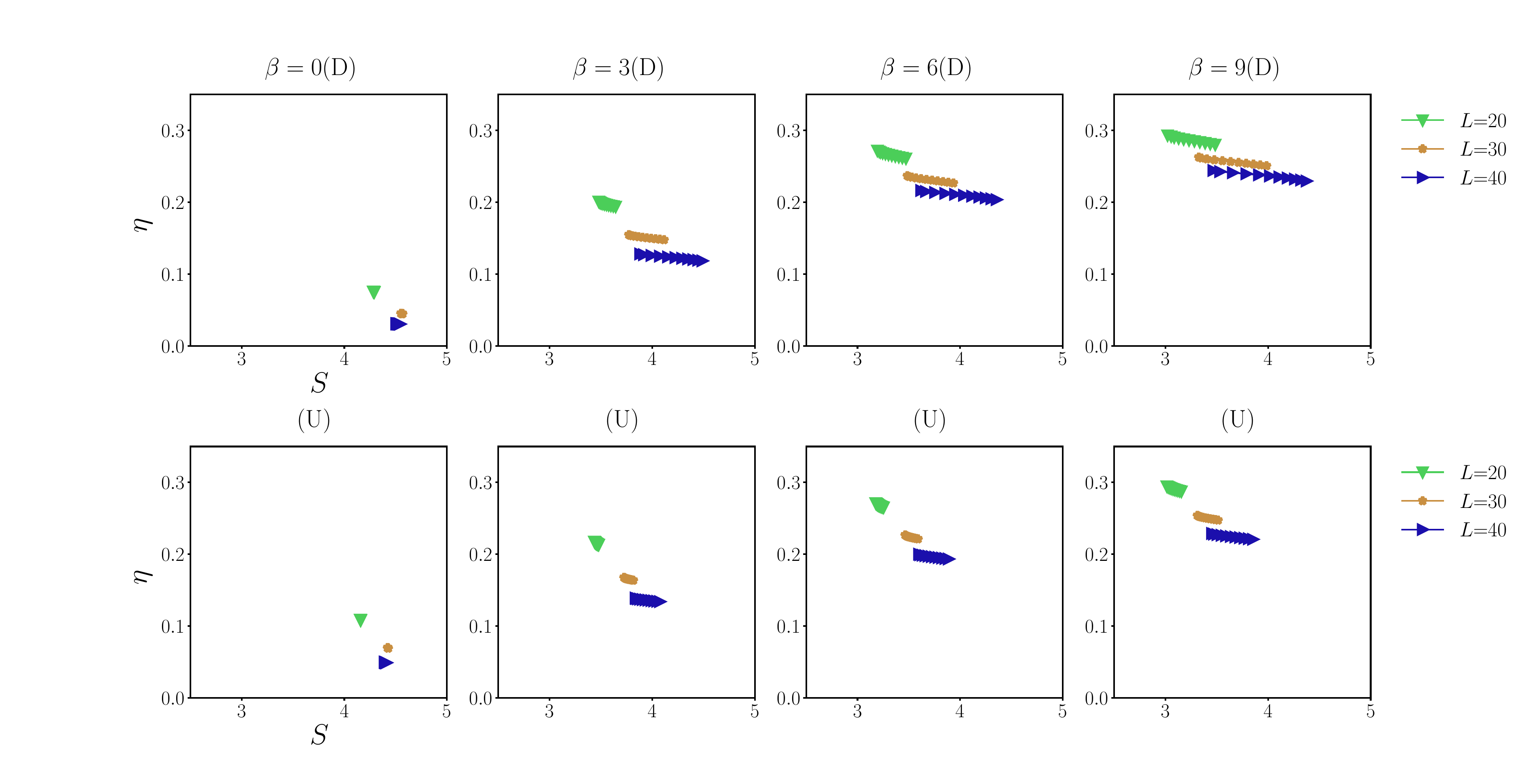} 
\caption{
The efficiency $\eta(h)$ vs the entropy production $S(h)$ for different $\beta=\{0,2,4,6\}$. The results are obtained by averaging over $1000,500,200$ independent runs of simulation with lattice sizes $L=20,30,40$, $c=1.0$ and $\kappa=1$. Top (bottom) panels show the results for the delta (uniform) energy distribution.  
}\label{f10}
\end{figure}

The strong effect of interactions on the efficiency and entropy can be mitigated by adjusting the link capacities according to the local populations as in Eq. \ref{Jstar} for $\kappa=1$. In Figs. (\ref{f7}) and (\ref{f8}) we report these quantities in terms of the parameters $\beta$ and $h$, respectively. It is now around $h=6$ that the qualitative behavior of the entropy with $\beta$ changes for the delta energy distribution, see panels (b1) of the figures. Compare it with the behavior of the uniform energy distribution in panels (b2). In addition, panels (a3) show that for $\beta$ smaller than about $4$ efficiency of the     
uniform case outperforms that of the delta distribution in a large range of the interaction strengths.

Figures (\ref{f9}) and (\ref{f10}) display the size dependence of the efficiency and entropy vs the parameters $\beta$ and $h$, respectively. Overall, the larger systems results in smaller efficiencies and generate more uncertainty in the return times than the smaller systems. In addition, we observe that the transition from a desirable regime (high $\eta$ and low $S$) to an undesirable one (low $\eta$ and high $S$) is sharper and occur at a larger $S$ when the initial energy budgets are the same (D) compared to the case of having a maximal energy variability (U). Moreover, the nonmonotonic behavior of $\eta(S)$ appears at larger values of interactions ($h$) when the initial energies are distributed uniformly.

\section{Conclusion}\label{S3}
We studied a model of interacting agents moving on a two-dimensional network using an exploration and return strategy which is governed by the available energy of the agents. We considered two distributions of initial energy budgets with very large and small energy variances but the same average energy. Energy is dissipated in the movement process with a rate that depends on the flow of particles in the links of the network. In summary, we observed that variability in initial energy helps to have in average a larger efficiency (distinct visited sites per consumed energy) when the return strategy dominates and a smaller entropy production (uncertainty in the return times) when exploration is preferred, specially in presence of strong interactions. 

This study provided another example of a movement process where in average a larger entropy production results in a smaller efficiency \cite{indaco-2021}. And, in the regime of strong interactions, the efficiency can have multiple values for a given level of uncertainty in the return times. In this paper, we worked with a different model of movements and used different definitions of efficiency and entropy production from the previous works, nevertheless, the qualitative behavior of these quantities remains the same. It would be interesting to investigate analytical and numerical solutions to the continuum  master equation in higher dimensions to see how much the results are affected by increasing the space dimension. 

The main assumption in this work was that available energy determines the importance of exploration and return in the movements. In addition, we assumed that interactions between the agents increases the rate of energy dissipation but not the travel times of the agents. In a more realistic model the flow of particles in a link can also affect the travel or waiting times along the links. This effect can have a large impact on the values of the efficiency and uncertainty in the return times.

\acknowledgments
This work was performed using the ALICE compute resources provided by Leiden University.

\bibliography{references}

\end{document}